\documentclass[12pt]{article}
\usepackage{epsfig,latexsym,graphicx}

\title{Coin-Moving Puzzles}
\author{%
  Erik D. Demaine%
    \thanks{Department of Computer Science, University of Waterloo, Waterloo,
            Ontario N2L 3G1, Canada, email: \{\texttt{eddemaine},
            \texttt{mldemaine}\}\texttt{@uwaterloo.ca}}
\and
  Martin L. Demaine%
    \footnotemark[1]
\and
  Helena A. Verrill%
    \thanks{Institut for Matematiske Fag, Universitetsparken 5,
            DK-2100 K\o benhavn, Denmark, email: \texttt{verrill@math.ku.dk}}}
\date{}

\advance \topmargin by -\headheight
\advance \topmargin by -\headsep
\evensidemargin \oddsidemargin
\let\latexcite=\cite
\def\cite{\nolinebreak\latexcite}
\let\latexref=\ref
\def\ref{\nolinebreak\latexref}

{\makeatletter \gdef\fps@figure{!htbp}}

\newtheorem{theorem}{Theorem}
\newtheorem{lemma}{Lemma}

\makeatletter
\def\GrabProofArgument[#1]{ (#1): \egroup\ignorespaces}
\def\proof{\noindent\textbf\bgroup Proof%
           \@ifnextchar[{\GrabProofArgument}{: \egroup\ignorespaces}}

\makeatother

{\makeatletter
 \gdef\xxx{\@ifnextchar[\xxx@lab\xxx@nolab}
 \long\gdef\xxx@lab[#1]#2{{\bf [\marginpar{xxx}{\sc #1}: #2]}}
 \long\gdef\xxx@nolab#1{{\bf [\marginpar{xxx}#1]}}
 \long\gdef\xxx@lab[#1]#2{}\long\gdef\xxx@nolab#1{}%
}

\def\captionfont{\sf\small}
\def\captionlabelfont{\bf\small}
{\makeatletter
 \global\let\old@makecaption\@makecaption
 \long\gdef\@makecaption#1#2{%
   \old@makecaption{\captionlabelfont #1}{\captionfont #2}}}

\def\span{\mathop{\rm span}}
\def\applymove{\,/\,}

\def\coinscale{0.75}

\let\setminus=-

\begin{document}
\maketitle

\begin{abstract}
We introduce a new family of one-player games, involving the movement of coins
from one configuration to another.  Moves are restricted so that a coin can be
placed only in a position that is adjacent to at least two other coins.  The
goal of this paper is to specify exactly which of these games are solvable.
By introducing the notion of a constant number of extra coins, we give tight
theorems characterizing solvable puzzles on the square grid and
equilateral-triangle grid.  These existence results are supplemented by
polynomial-time algorithms for finding a solution.
%
%---Old:
% We analyze a family of one-player games, involving the movement of coins from
% one configuration to another.  Moves are restricted so that a coin can be
% placed only in a position that is adjacent to at least two other coins.  The
% goal of this paper is to specify exactly which of these games are solvable.  We
% solve this problem with a polynomial-time algorithm for two boards, the
% equilateral-triangle and square grids, and consider the problem on a general
% graph.
\end{abstract}

\section{Introduction}

\begin{figure}
\centerline{\includegraphics[scale=\coinscale]{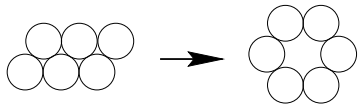}}
\caption{\label{rhomboid}
  Re-arrange the rhombus into the circle using three slides,
  such that each coin is slid to a position adjacent to two other coins.}
\end{figure}

Consider a configuration of coins such as the one on the left of Figure
\ref{rhomboid}.  The player is allowed to move any coin to a position that is
determined rigidly by incidences to other coins.  In other words, a coin can be
moved to any position adjacent to at least two other coins.  The puzzle or
1-player game is to reach the configuration on the right of Figure
\ref{rhomboid} by a sequence of such moves.  This particular puzzle is most
interesting when each move is restricted to \emph{slide} a coin in the plane
without overlapping other coins.

\begin{figure}
\centerline{\includegraphics[scale=\coinscale]{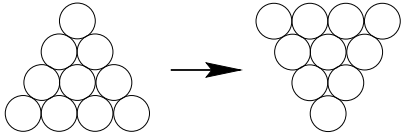}}
\caption{\label{upside-down}
  Turn the pyramid upside-down in three moves,
  such that each coin is moved to a position adjacent to two other coins.}
\end{figure}

\begin{figure}
\centerline{\includegraphics[scale=\coinscale]{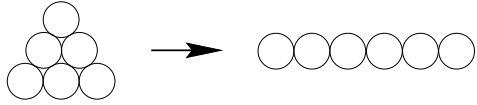}}
\caption{\label{original helena}
  Re-arrange the pyramid into a line in seven moves,
  such that each coin is moved to a position adjacent to two other coins.}
\end{figure}

This puzzle is described in Gardner's Mathematical Games article on Penny
Puzzles \cite{Gardner-1975-penny}, in \emph{Winning Ways}
\cite{Berlekamp-Conway-Guy-1982-coins}, in \emph{Tokyo Puzzles}
\cite{Fujimura-1978-coins}, in \emph{Moscow Puzzles}
\cite{Kordemsky-1972-coins}, and in \emph{The Penguin Book of Curious and
Interesting Puzzles} \cite{Wells-1992-coins}.  Langman \cite{Langman-1951}
shows all 24 ways to solve the puzzle in three moves.  Another classic puzzle
of this sort \cite{Bolt-1984-pennies, Fujimura-1978-coins, Gardner-1975-penny,
Wells-1992-coins} is shown in Figure \ref{upside-down}.  A final classic puzzle
that originally motivated our work is shown in Figure \ref{original helena};
its source is unknown.  Other related puzzles are presented by Dudeney
\cite{Dudeney-1967-pennies}, Fujimura \cite{Fujimura-1978-coins}, and Brooke
\cite{Brooke-1963}.
% Brooke covers Figure \ref{rhomboid} and some related puzzles,
% but not Figure \ref{upside-down}.

\begin{figure}
\centerline{\includegraphics[scale=\coinscale]{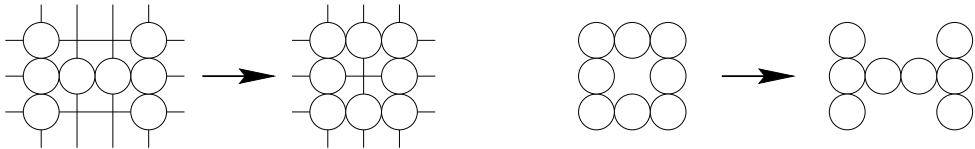}}
\caption{\label{HOH}
  Re-arrange the H into the O in four moves while staying on the square grid
  (and always moving adjacent to two other coins),
  and return to the H in six moves using both the equilateral-triangle and
  square grids.}
\end{figure}

The puzzles above always move the centers of coins to vertices of the
equilateral-triangle grid.  Another type of puzzle is to move coins on the
square grid, which appears less often in the literature but has significantly
more structure and can be more difficult.  The only published example we are
aware of is given by Langman \cite{Langman-1953}, which is also described by
Brooke \cite{Brooke-1963}, Bolt \cite{Bolt-1991-coins}, and Wells
\cite{Wells-1992-coins}; see Figure \ref{HOH}.  The first puzzle (H $\to$ O) is
solvable on the square grid, and the second puzzle (O $\to$ H) can only be
solved by a combination of the two grids.

\begin{figure}
\centerline{\includegraphics[scale=\coinscale]{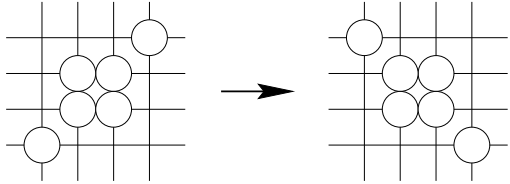}}
\caption{\label{diagonal}
  A difficult puzzle on the square grid.
  The optimal solution uses 18 moves,
  each of which places a coin adjacent to two others.}
\end{figure}

\begin{figure}
\centerline{\includegraphics[scale=\coinscale]{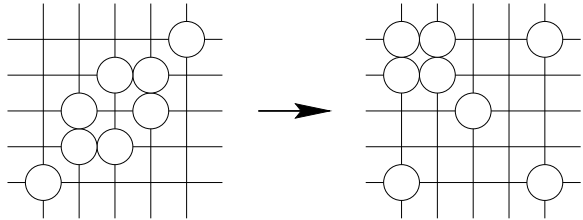}}
\caption{\label{spindle}
  Another puzzle on the square grid.
  The optimal solution uses 24 moves,
  each of which places a coin adjacent to two others.}
\end{figure}

\begin{figure}
\centerline{\includegraphics[scale=\coinscale]{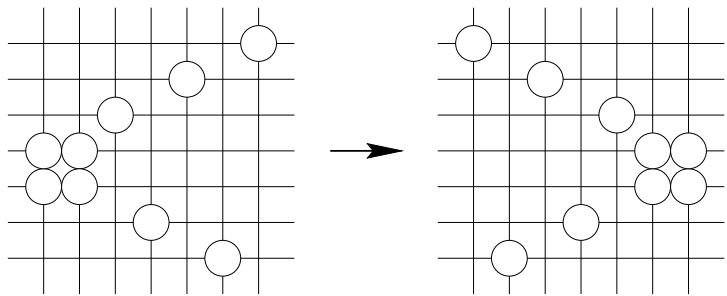}}
\caption{\label{vee}
  Another puzzle on the square grid with the same rules.}
  %Solvable in 21 moves, each of which places a coin adjacent to two others.}
\end{figure}

\begin{figure}
\centerline{\includegraphics[scale=\coinscale]{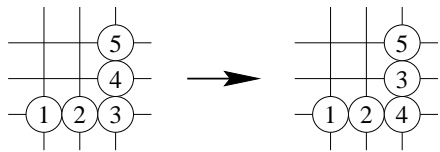}}
\caption{\label{swap5}
  A puzzle on the square grid involving labeled coins.
  Solvable in eleven moves, each of which places a coin adjacent to two
  others; see Figure \protect\ref{swap5 solution}.}
\end{figure}

\begin{figure}
\centerline{\includegraphics[scale=0.8]{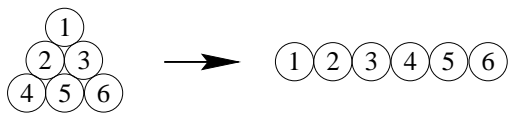}}
\caption{\label{tri2oline}
  A puzzle on the equilateral-triangle grid involving labeled coins.
  Solvable in eight moves, each of which places a coin adjacent to two others.}
\end{figure}

In this paper we study generalizations of these types of puzzles, in which
coins are moved on some grid to positions adjacent to at least two other coins.
Specifically, we address the basic algorithmic problem: is it possible to solve
a puzzle with given start and finish configurations, and if so, find a
sequence of moves.  Surprisingly, we show that this problem has a
polynomial-time solution in many cases.  Our goal in this pursuit is to gain a
better understanding of what is possible by these motions, and as a result to
design new and interesting puzzles.  For example, one puzzle we have designed
is shown in Figure \ref{diagonal}.  We recommend the reader try this difficult
puzzle before reading Section \ref{Re-orienting L's} which shows how to solve
it.  Figures \ref{spindle}--\ref{tri2oline} show a few of the other puzzles we
have designed.  The last two puzzles involve labeled coins.

\xxx{A variation on `diagonal' is where the extra coins are labeled differently
from the others.  How much harder does this make the puzzle?  Try it and
mention it.}

This paper studies two grids in particular: the equilateral-triangle grid, and
the square grid.  It turns out that the triangular grid has a relatively simple
structure, and nearly all puzzles are solvable.  An exact, efficient
characterization of solvable puzzles is presented in Section \ref{Triangular
Grid}.  The square grid has a more complicated structure, requiring us to
introduce the notion of ``extra coins'' to give a partial characterization of
solvable puzzles.  This result is described in Section \ref{Square Grid} after
some general tools for analysis are developed in Section \ref{General Tools}.

Before we begin, the next section defines a general graph model of the puzzles
under consideration.

%We begin in Section \ref{General Theory} with a general graph model of
%token-moving puzzles.  Section \ref{Square Grid} presents our main result about
%the square grid.  In Section \ref{Triangular Grid} we describe our result about
%the equilateral-triangle grid.  We conclude in Section \ref{Conclusion} with
%several open problems.

%\pagebreak
\section{Model}
\label{Model}

%This section provides some general concepts that we will find useful for
%analyzing the games.  We begin in Section \ref{Problem Statement} with a formal
%problem statement and definitions of terminology.  Section \ref{Span} presents
%the notion of the span of a configuration, which essentially specifies the
%reachable positions given an unlimited supply of extra coins.
%
%\subsection{Problem Statement}
%\label{Problem Statement}

We begin by defining ``token-moving'' and ``coin-moving'' puzzles
and related concepts.
%We begin with some basic concepts generic to token-moving and coin-moving
%puzzles.
The \emph{tokens} form a finite multiset $T$.  We normally think of
tokens as unlabeled, modeled by all elements of $T$ being equal, but another
possibility is to color tokens into more than one equivalence class (as in
Figure \ref{tri2oline}).  A \emph{board} is any simple undirected graph
$G=(P,E)$, possibly infinite, whose vertices are called \emph{positions}.  A
\emph{configuration} is a placement of the tokens onto distinct positions on
the board, i.e., a one-to-one mapping $C : T \to P$.  We will often associate a
configuration $C$ with its image, that is, the set of positions \emph{occupied}
by tokens.

A \emph{move} from a configuration $C$ changes the position of a single token
$t$ to an unoccupied position $p$, resulting in a new configuration.  This move
is denoted $t \mapsto p$, and the resulting configuration is denoted $C
\applymove t \mapsto p$.  We stress that moves are not required to ``slide''
the token while avoiding other tokens (like the puzzle in Figure
\ref{rhomboid}); the token can be picked up and placed in any unoccupied
position.

The \emph{configuration space} (or \emph{game graph}) is the directed graph
whose vertices are configurations and whose edges correspond to feasible moves.
A typical \emph{token-moving puzzle} asks for a sequence of moves to reach one
configuration from another, i.e., for a path between two vertices in the
configuration space, subject to some constraints.  A \emph{coin-moving puzzle}
is a geometric instance of a token-moving puzzle, in which tokens are
represented by \emph{coins}---constant-radius disks in the plane, and
constant-radius hyperballs in general---and the board is some lattice in the
same dimension.  If a token-moving or coin-moving puzzle with source
configuration $A$ and destination configuration $B$ is solvable, we say that
$A$ can be \emph{re-arranged} into $B$, and that $B$ is \emph{reachable} from
$A$.  This is equivalent to the existence of a directed path from $A$ to $B$
in the configuration space.

This paper addresses the natural question of what puzzles are solvable, subject
to the following constraint on moves which makes the problem interesting.  A
move $t \mapsto p$ is \emph{$d$-adjacent} if the new position $p$ is adjacent
to at least $d$ tokens other than the moved token $t$.  (Throughout,
\emph{adjacency} refers to the board graph $G$.)  This constraint is
particularly meaningful for $d$-dimensional coin-moving puzzles, because then a
move is easy to ``perform exactly'' without any underlying lattice: the new
position $p$ is determined rigidly by the $d$ coin adjacencies
(sphere tangencies).

The \emph{$d$-adjacency configuration space} is the subgraph of the
configuration space in which moves are restricted to be $d$-adjacent.
Studying connectivity in this graph is equivalent to studying solvable puzzles;
for example, if the graph is strongly connected, then all puzzles are solvable.

Here we explore solvable puzzles on two boards, the equilateral-triangle grid
and the square grid.  Because these puzzles are two-dimensional, in the context
of this paper we call a move \emph{valid} if it is 2-adjacent, and a position a
\emph{valid destination} if it is unoccupied and adjacent to at least two
occupied positions.  Thus a valid move involves transferring a token from some
source position to a valid destination position.  When the context is clear, we
will refer to a valid move just by ``move.''  A move is \emph{reversible} if
the source position is also a valid destination.

\section{Triangular Grid}
\label{Triangular Grid}

This section studies the equilateral-triangle grid, where most puzzles are
solvable.  To state our result, we need a simple definition.  Associated with
any configuration is the subgraph of the board induced by the occupied
positions.  In particular, a \emph{connected component} of a configuration is a
connected component in this induced subgraph.

\begin{theorem} \label{triangular grid}
On the triangular grid with the 2-adjacency restriction and unlabeled coins,
configuration $A$ can be re-arranged into a different configuration $B$
precisely if $A$ has a valid move, the number of coins in $A$ and $B$ match,
and at least one of four conditions holds:

  \begin{enumerate}
  \item $B$ contains three coins that are mutually adjacent (a triangle).
  \item $B$ has a connected component with at least four coins.
  \item $B$ has a connected component with at least three coins
        and another connected component with at least two coins.
  \item There is a single move from $A$ to $B$.
  \end{enumerate}

\noindent
The same result holds for labeled coins, except when there are exactly three
coins in the puzzle, in which case the labelings and movements are controlled
by the vertex 3-coloring of the triangular grid.

Furthermore, there is a polynomial-time algorithm to find a re-arrangement from
$A$ to $B$ if one exists.  Specifically, let
$n$ denote the number of coins and $d$ denote the maximum
distance between two coins in $A$ or $B$.
Then a solution with $O(n d)$ moves can be found in $O(n d)$ time.
\end{theorem}

The rest of this section is devoted to the proof of this theorem.
We begin in the next subsection by proving necessity of the conditions: if a
puzzle is solvable, then one of the conditions holds.  Then in the following
subsection we prove sufficiency of the conditions.

\subsection{Necessity}

Of course, it is necessary for $A$ to have a valid move and for $A$
and $B$ to have the same number of coins.  Necessity of at least one of the
four conditions is also not difficult to show, because Conditions 1--3 are so
broad, encompassing most possibilities for configuration $B$.

\begin{figure}
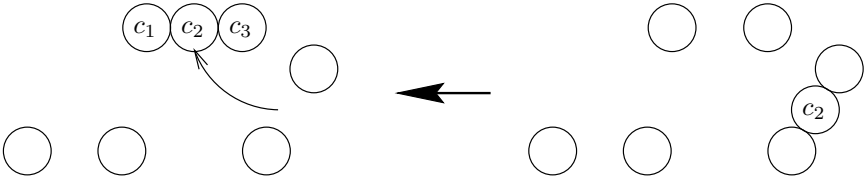

\centerline{\input condition4.pstex_t}
\caption{\label{condition4}
  Reverse-moving a configuration $B$ that does not satisfy any of
  Conditions 1--3.}
\end{figure}

Suppose that a solvable puzzle does not satisfy any of Conditions 1--3, as in
Figure \ref{condition4}.  We prove that it must satisfy Condition 4, by
considering play backwards from the goal configuration $B$.  Specifically, a
\emph{reverse move} takes a coin currently adjacent to at least two others, and
moves it to any other location.  Because the puzzle is solvable, some coin in
configuration $B$ must be reverse-movable, i.e., must have at least two coins
adjacent to it.  Thus, some connected component of $B$ has at least three
coins.  Because Condition 2 does not hold, this connected component has exactly
three coins.  Because Condition 1 does not hold, these three coins are not
connected in a triangle.  Because Condition 3 does not hold, every other
component has exactly one coin.

Hence, one component of $B$ is a path of exactly three coins, say $c_1, c_2,
c_3$, and every other component of $B$ has exactly one coin, as in the left of
Figure \ref{condition4}.  Certainly at this moment $c_2$ is the only
reverse-movable coin.  We claim that after a sequence of reverse moves, $c_2$
will continue to be the only reverse-movable coin.  If we removed $c_2$, then
every coin would be adjacent to no others.  Thus, if we reverse move $c_2$
somewhere, then every other coin would be adjacent to at most one other
($c_2$).  Hence, it remains that only $c_2$ can be reverse moved.

Therefore, if we can reach $A$ from $B$ via reverse moves, we can do so in a
single reverse move of $c_2$ directly to where it occurs in $A$.  Thus
Condition 4 holds, as desired.

\subsection{Sufficiency}

Next we prove the more difficult direction: provided one of Conditions 1--3
hold, there is a re-arrangement from $A$ to $B$.  (This fact is obvious when
Condition 4 holds.)  All three cases will follow a common outline: we first
form a triangle (Section \ref{Getting Started}), then maneuver this triangle
(Section \ref{Triangle Maneuvering}) to transport all other coins (Section
\ref{Transportation}), and finally we place the three triangle coins
appropriately depending on the case (Section \ref{Finale}).

\subsubsection{Getting Started}
\label{Getting Started}

\begin{figure}
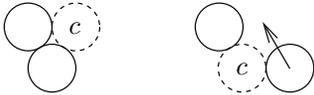

\centerline{\input move_tri.pstex_t}
\caption{\label{move tri}
  Two types of valid destinations for a coin $c$.
  In the latter case, we show a move to form a triangle.}
\end{figure}

It is quite simple to make some triangle of coins.  By assumption, there is a
valid move from configuration $A$.  The destination of this move can have two
basic forms, as shown in Figure \ref{move tri}.  Either the move forms a
triangle, as desired, or the move forms a path of three coins.  In the latter
case, if there is not a triangle already with a different triple of coins, a
triangle can be formed by one more move as shown in the right of the figure.

\begin{figure}
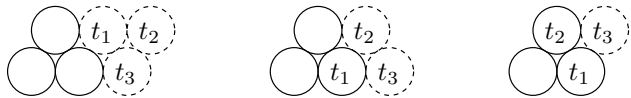

\centerline{\input bootstrap_tri.pstex_t}
\caption{\label{bootstrap tri}
  The three cases of building a triangle $\{t_1, t_2, t_3\}$ out of an
  existing triangle, depending on how many coins the two triangles share.
  {From} left to right, zero, one, and two coins of overlap.}
\end{figure}

This triangle $T_0$ suffices for unlabeled coin puzzles.  However, for labeled
coin puzzles, we cannot use just any three coins in the triangle; we need a
particular three, depending on $B$.  For example, if $B$ satisfies Condition 1,
then the coins forming the triangle in $B$ are the coins we would like in the
triangle for maneuvering.  To achieve this, we ``bootstrap'' the
triangle $T_0$ formed above, using this triangle with the incorrect coins to
form another triangle with the correct coins.  Specifically, if we desire a
triangle using coins $t_1$, $t_2$, and $t_2$, then we move each coin
in the difference $\{t_1, t_2, t_3\} \setminus T_0$ to be adjacent to
appropriate coins in $T_0$.  There are three cases, shown in Figure
\ref{bootstrap tri}, depending on how many coins are in the difference.  If
ever we attempt to move a coin to an already occupied destination, we first
move the coin located at that destination to any other valid destination.

\subsubsection{Triangle Maneuvering}
\label{Triangle Maneuvering}

Consider a triangle of coins $t_1$, $t_2$, and $t_3$.  The possible positions
of this triangle on the triangular grid are in one-to-one correspondence with
their centers, which are vertices of the dual hexagonal grid.  Moving one coin
(say $t_1$) to be adjacent to and on the other side of the others ($t_2$ and
$t_3$) corresponds to moving the center of the triangle to one of the three
neighboring centers on the hexagonal grid.  Thus, without any other coins on
the board, the triangle can be moved to any position by following a path in the
hexagonal grid.

\begin{figure}
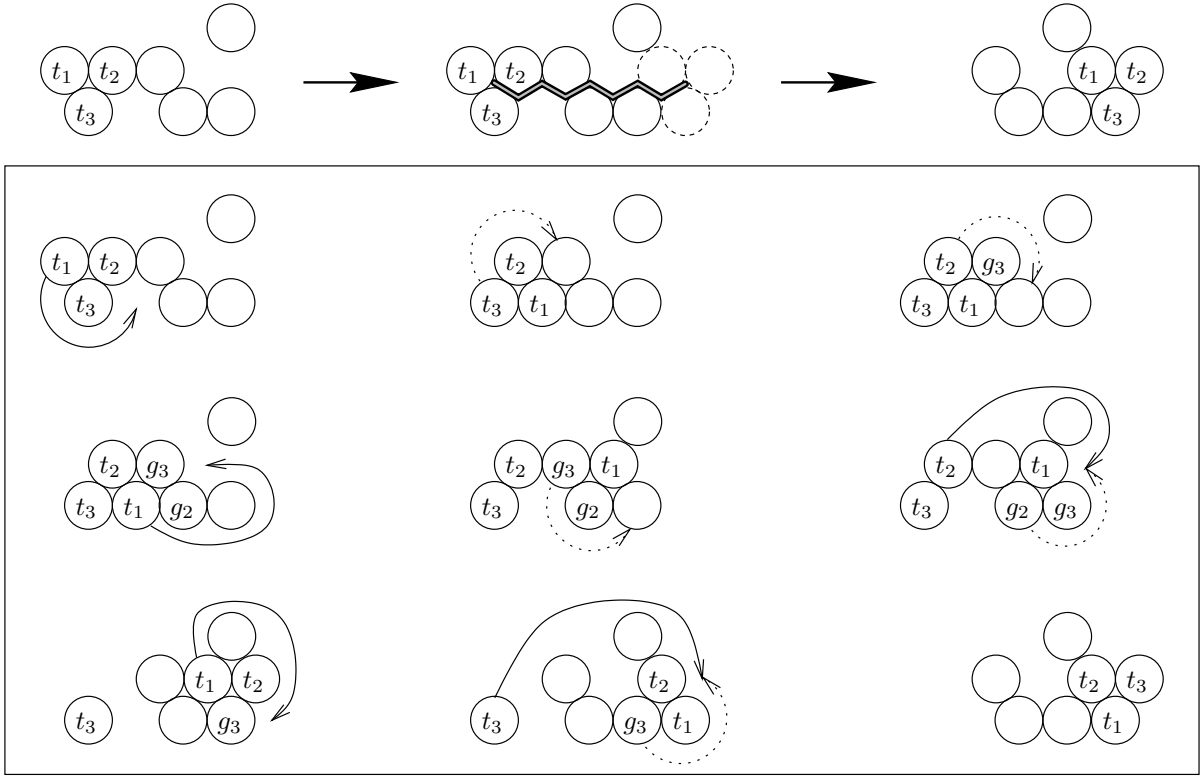

\centerline{\input maneuvering_triangle.pstex_t}
\caption{\label{maneuvering triangle}
  An example of triangle maneuvering.  Dotted arrows denote conceptual moves,
  and solid arrows denote actual moves.}
\end{figure}

This approach can be modified to apply when there are additional obstacle
coins; see Figure \ref{maneuvering triangle} for an example.  Conceptually we
always move one of the triangle coins, say $t_i$, in order to move the center
of the triangle to an adjacent vertex of the hexagonal grid.  But if the move
of $t_i$ is impossible because the destination is already occupied by another
coin $g_i$, then in fact we do not make any move.  There will be a triangle in
the desired position now, but it will not consist of the usual three coins
($t_1$, $t_2$, and $t_3$); instead, $t_i$ will be replaced by the ``ghost
coin'' $g_i$.  Such a triangle suffices for our purposes of transportation
described in Section \ref{Transportation}.  One final detail is how the ghost
coins behave: if we later need to move a ghost coin $g_i$, we instead move the
original (unmoved) coin $t_i$.  Thus ghost coins are never moved; only $t_1$,
$t_2$, and $t_3$ are moved during triangle maneuvering (even if coins are
labeled).

\subsubsection{Transportation}
\label{Transportation}

Triangle maneuvering makes it easy to \emph{transport} any other coin to any
desired location.  Specifically, suppose we want to move coin $c \notin \{t_1,
t_2, t_3\}$ to destination position $d$.  If $d$ is already occupied by another
coin $c'$, we first move $c'$ to an arbitrary valid destination; there is at
least one because the triangle can be maneuvered.  Now we maneuver the triangle
so that the (potentially ghost) triangle has two coins adjacent to $d$, so that
the third coin is not on $d$, and so that the triangle does not overlap $c$.
This is easily arranged by examining the location of $c$ and setting the
destination of the triangle appropriately.  For example, if $c$ is within
distance two of $d$, then there are four positions for the triangle that are
adjacent to $d$ and do not overlap $c$; otherwise,
the triangle can be placed in any of the six positions adjacent to $d$.
Finally, because $d$ is now a valid destination---it is adjacent to two coins
in the triangle---we can move $c$ to $d$.

\begin{figure}
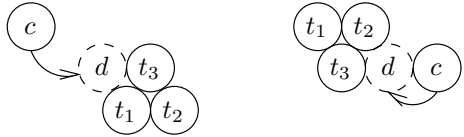

\centerline{\input transportation_tri.pstex_t}
\caption{\label{transportation tri}
  Transporting coin $c$ to destination $d$ using triangle $\{t_1, t_2, t_3\}$.
  In both cases, we choose the location of the triangle
  so that it does not overlap $c$.}
\end{figure}

By the properties of triangle maneuvering, this transportation process even
preserves coin labels: the only actual coins moved are $t_1$, $t_2$, $t_3$,
$c$, and possibly a coin at $d$.  But any coin at position $d$ must not have
already been in its desired position, because $d$ is $c$'s desired position.
Thus, applying the transportation process to every coin except $t_1$, $t_2$,
and $t_3$ places all coins except these three in their desired locations.

\subsubsection{Finale}
\label{Finale}

Once transportation is complete, it only remains to place the triangle coins
$t_1$, $t_2$, and $t_3$ in their desired locations.  By the bootstrapping in
Section \ref{Getting Started}, we are able to choose the unplaced coins $\{t_1,
t_2, t_3\}$ however we like.  This property will be exploited differently
in the three cases.
% For the moment we ignore the labeling of $t_1$, $t_2$, and $t_3$.

\paragraph{Property 1.}
If there is a triangle in $B$, then we choose these three coins as the unplaced
coins $t_1$, $t_2$, and $t_3$, and use them to transport all other coins.  Then
we maneuver the triangle $\{t_1, t_2, t_3\}$ exactly where it appears in $B$.
Because all other coins have been moved to their proper location, in this
position the triangle will not have any ghost coins.

\begin{figure}
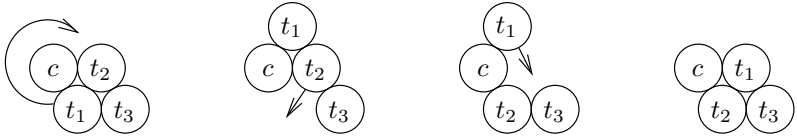

\centerline{\input relabel_tri.pstex_t}
\caption{\label{relabel tri}
  Swapping coins $t_1$ and $t_2$ in a triangle, using an adjacent coin $c$.}
\end{figure}

However, it may be that the coins $\{t_1, t_2, t_3\}$ are labeled incorrectly
among themselves, compared to $B$.  Assuming there are more than three coins in
the puzzle, this problem can be repaired as follows.  We maneuver the triangle
so that it does not overlap any other coins but is adjacent to at least one
coin $c$; for example, there is such a position for the triangle just outside
the smallest enclosing hexagon of the other coins.  Refer to Figure
\ref{relabel tri}.  Now two coins of the triangle, say $t_1$ and $t_2$, are
adjacent to three other coins each: each other, $t_3$, and $c$.  Thus we can
move $t_1$ to any other valid destination, and then move $t_2$ or $t_3$ to
replace it.  Afterwards we can move $t_1$ to take the place of $t_2$ or $t_3$,
whichever moved.  This procedure swaps $t_1$ and either $t_2$ or $t_3$.  By
suitable application, we can achieve any permutation of $\{t_1, t_2, t_3\}$,
and thereby achieve the desired labeling of the triangle.

\begin{figure}
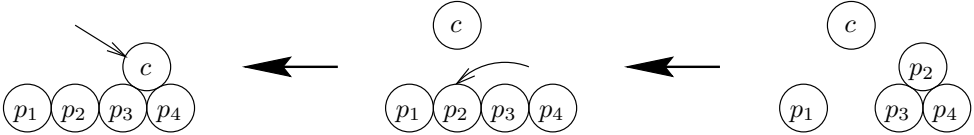

\centerline{\input finale_tri_2.pstex_t}
\caption{\label{finale tri 2}
  Reverse moving a configuration $B$ with Property 2 into a
  configuration with Property 1.}
\end{figure}

\paragraph{Property 2 but not Property 1.}
Refer to Figure \ref{finale tri 2}.  If there is not a triangle in $B$, but
there is a connected component of $B$ with at least four coins, then there is a
path in $B$ of length four, $(p_1, p_2, p_3, p_4)$.  From {B} we reverse move
$p_2$ so that it is adjacent to $p_3$ and $p_4$.  If this position is already
occupied by a coin $c$, we first reverse move $c$ to any other unoccupied
position.  Now $p_2$, $p_3$, and $p_4$ are mutually adjacent, so we have a
new destination configuration $B'$
with Property 1.  As described above, we can re-arrange $A$ into $B'$.
Then we undo our reverse moves: move $p_2$ back adjacent to $p_1$ and $p_3$,
and move $c$ back adjacent to $p_3$ and $p_4$.  This procedure re-arranges
$A$ into $B$.

\begin{figure}
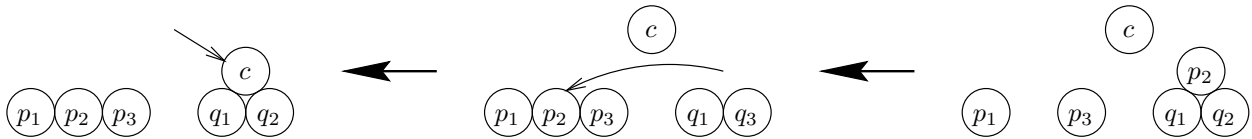

\centerline{\input finale_tri_3.pstex_t}
\caption{\label{finale tri 3}
  Reverse moving a configuration $B$ with Property 3 into a
  configuration with Property 1.}
\end{figure}

\paragraph{Property 3 but not Property 1.}
This case is similar to the previous one; refer to Figure \ref{finale tri 3}.
There must be a path in $B$ of length three, $(p_1, p_2, p_3)$, as well as a
pair of adjacent coins, $(q_1, q_2)$, in different connected components of $B$.
If both positions adjacent to both $q_1$ and $q_2$ are already occupied, we
first reverse move one such coin (call it $c$) to an arbitrary unoccupied
position.  This frees up a position adjacent to $q_1$ and $q_2$, to which we
reverse move $p_2$.  Now $\{q_1, q_2, p_2\}$ form a triangle, so Property 1
holds, and we can reach this new configuration $B'$ from $A$.  Then we undo our
reverse moves: move $p_2$ back adjacent to $p_1$ and $p_3$, and move $c$ back
adjacent to $q_1$ and $q_2$.  This procedure re-arranges $A$ into $B$.

This concludes the proof of Theorem \ref{triangular grid}.

%--- This is covered in the conclusion instead.
% \subsection{Conjectures}

% An interesting question is 

% \begin{conjecture}
% On the triangular grid with the 2-adjacency restriction and unlabeled coins, if
% there is a re-arrangement from $A$ to $B$ when the coins are unlabeled, then
% there is also a re-arrangement consisting of a sequence of \emph{slides}.
% \end{conjecture}

%--- Old stuff about span:
% Coin-moving puzzles have an entirely different structure with the triangular
% grid.  In particular, we have the following characterization of span:

% \begin{lemma}
% For a configuration $C$ on the triangular grid, if $C$ has a valid destination
% position, then $\span C$ is the entire triangular grid; otherwise, $\span C =
% C$.
% \end{lemma}

\section{General Tools}
\label{General Tools}

In this section we develop some general lemmas about token-moving puzzles.
Although we only use these tools for the square grid, in Section \ref{Square
Grid}, they apply to arbitrary boards and may be of more general use.

\subsection{Picking Up and Dropping Tokens}
\label{Picking Up and Dropping Tokens}

First we observe that additional tokens cannot ``get in the way'':
%First we observe that, for the purpose of showing connectivity of the
%configuration space, tokens can be removed freely because this modification
%only makes the puzzle more difficult.

\begin{lemma} \label{emulation}
If a token-moving puzzle is solvable, then it remains solvable if we add an
additional token with an unspecified destination, provided tokens are
unlabeled.  This result also holds if all moves must be reversible.
\end{lemma}

\begin{proof}
A move can be blocked by an extra token $e$ at position $p$ because $p$ is
occupied and hence an invalid destination.  But if ever we encounter such a
move of a token $t$ to position $p$, we can just ignore the move, and swap the
roles of $e$ and $t$: treat $e$ as the moved version of $t$, and treat $t$
as an extra token replacing $e$.  Thus, any sequence of moves in the original
puzzle can be emulated by an equivalent sequence of moves in the augmented
puzzle.  We are not introducing any new moves, only removing existing moves,
so all moves remain reversible if they were originally.
\end{proof}

This proof leads to a technique for emulating a more powerful model for solving
puzzles.  In addition to moving coins as in the normal model, we can
conceptually \emph{pick up} (remove) a token, and later \emph{drop} (add) it
onto any valid destination.  At any moment we can have any number of tokens
picked up.  While a token $t$ is conceptually picked up, we emulate any moves
to its actual position $p$ as in the proof of Lemma \ref{emulation}: if we
attempt to move another token $t'$ onto position $p$, we instead reverse the
roles of $t$ and $t'$.  To drop a token onto a desired position $p$, we simply
move the actual token to position $p$ if it is not there already.

Of course, this process may permute the tokens.  Nonetheless we will find this
approach useful for puzzles with labeled tokens.

One might instead consider the emulation method used implicitly in
Section \ref{Triangle Maneuvering} for triangular maneuvering: move original
coins instead of ghost coins.  This approach has the advantage that it
preserves the labels of the coins.  Unfortunately, the approach makes it
difficult to preserve reversibility as in Lemma \ref{emulation}, and so is
insufficient for our purposes here.

\subsection{Span}
\label{Span}

The \emph{span} of a configuration $C$ is defined recursively as follows.  Let
$d_1, \dots, d_m$ be the set of valid destinations for moves in $C$.
%--- Valid destination is now defined, so we don't need this:
%For example, under the $k$-adjacency restriction, these
%are all unoccupied positions that have at least $k$ occupied neighbors.
%--- This is a ``tighter'' definition of span, but is somehow less intuitive:
%Let $t_1 \mapsto d_1, \dots, t_m \mapsto d_m$ be the valid moves from
%$C$, subject to any restrictions imposed on the puzzle.
If $m=0$, the span of $C$ is just $C$ itself.  Otherwise, it is the span of
another configuration $C'$, defined to be $C$ with additional tokens at
positions $d_1, \dots, d_m$.  If this process never terminates, the span is
defined to be the limit, which exists because it is a countable union of finite
sets.

\begin{figure}
\centerline{\includegraphics[scale=\coinscale]{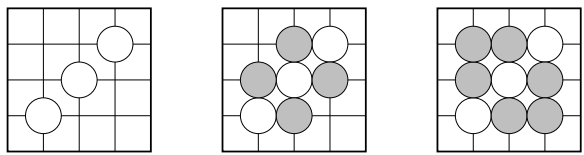}}
\caption{\label{3x3 diagonal span}
  In this example, the span is the smallest rectangle enclosing the
  configuration.}
\end{figure}

The span of a configuration lists all the positions we could hope to reach, or
more precisely, the positions we could reach if we had an unlimited number of
extra tokens that we could drop.  In particular, we have the following:

\begin{lemma} \label{span monotone}
If configuration $A$ can be re-arranged into configuration $B$, then $\span A
\supseteq B$ and thus $\span A \supseteq \span B$.
\end{lemma}

In other words, valid moves can never cause the span of the current
configuration to increase.  Thus the most connected we could hope the
configuration space to be is the converse of Lemma \ref{span monotone}: for
every pair of configurations with $\span A \supseteq \span B$, $A$ can be
re-arranged into $B$.  In words, we want that every configuration $A$ can be
re-arranged into any configuration $B$ with the \emph{same or smaller span}.

We call a configuration \emph{span-minimal} if the removal of any of its tokens
reduces the span.  Span-minimal configurations are essentially the ``skeleta''
that keep configurations with the same span reachable.
%
%--- The following is a funny way to put the result... this result is about
%    the number of coins past span minimality, which is different from ``extra
%    coins.''  For example, in triangular grid this is meaningless, though the
%    lemma still holds.
% 
% \subsection{Zero Extra Tokens Can Be Insufficient}
% 
% At this point it is easy to justify having at least one extra token
% in the worst case for any board:
%
One general property of span-minimal configurations is the following:

\begin{lemma} \label{span-minimal fragile}
If a configuration is span-minimal, any move will reduce the span.
\end{lemma}

\begin{proof}
Suppose to the contrary that there is a move $t \mapsto p$ that does not reduce
the span of a span-minimal configuration $C$.  In particular, $p$ must be a
valid destination position in the subconfiguration $C-t$, because $t$ does not
count in the $d$-adjacency restriction.  Hence, adding a new token at position
$p$ to the configuration $C-t$ has no effect on the span of $C$.  But this
two-step process of removing token $t$ and adding a token at position $p$ is
equivalent to moving $t$ to $p$, so $\span (C \applymove t \mapsto p) = \span
(C-t)$.  But we assumed that $\span (C \applymove t \mapsto p) = \span C$, and
hence $\span C = \span (C-t)$, contradicting that $C$ is span-minimal.
\end{proof}

Under the 2-adjacency restriction, a \emph{chain} is a sequence of tokens with
the property that the distance (in the board graph $G$) between two successive
tokens is at most $2$.  We will use chains as basic ``units'' for creating a
desired span.

%----- Probably not useful for general $k$:
% Under the $k$-adjacency restriction, a \emph{chain} is a sequence of occupied
% positions with the property that the distance between two consecutive positions
% (in the board graph $G$) is at most $k$.  We will use chains as basic ``units''
% for creating a desired span.

%--- Old, and pretty useless:
%For each type of board, the basic approach of our analysis is as follows.
%First, we give some basic properties of the span.  Then we determine the
%meaning of the span in terms of transferability.

Notice that the notion of span is useless for the already analyzed triangular
grid: provided there is a valid move, the span of any configuration is the
entire grid.  Thus, for the triangular grid, a configuration is span-minimal
precisely if it has no valid moves.  For the square grid, however, the notion
of span and span minimality is crucial.

\subsection{Extra Tokens}
\label{Extra Tokens}

\begin{figure}
\centerline{\includegraphics[scale=0.55]{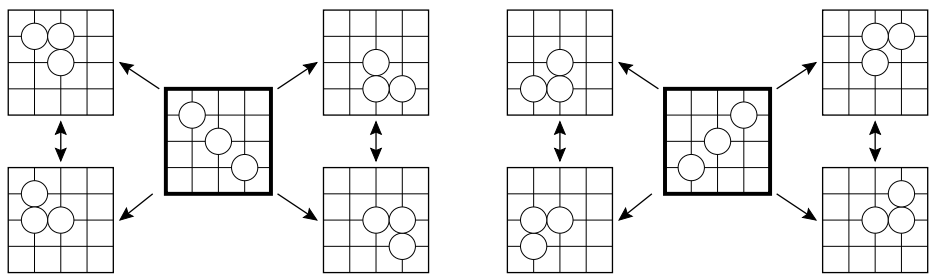}}
\caption{\label{square grid 0 extra}
  Subgraph of configuration space reachable from full-span configurations
  (outlined in bold) with no extra coins.}
\end{figure}

As described in the previous section, we can only re-arrange configurations
into configurations with the same or smaller span.  Unfortunately, the converse
is not true.  Indeed, the key problem situations are span-minimal
configurations; by Lemma \ref{span-minimal fragile}, such configurations
immediately lose span when we try to move them.  Hence, any two distinct
span-minimal configurations with the same span cannot reach each other.  An
example on the square grid is that the two opposite diagonals of a square are
unreachable from each other, as shown in Figure \ref{square grid 0 extra}.

\begin{figure}
\centerline{\includegraphics[scale=0.55]{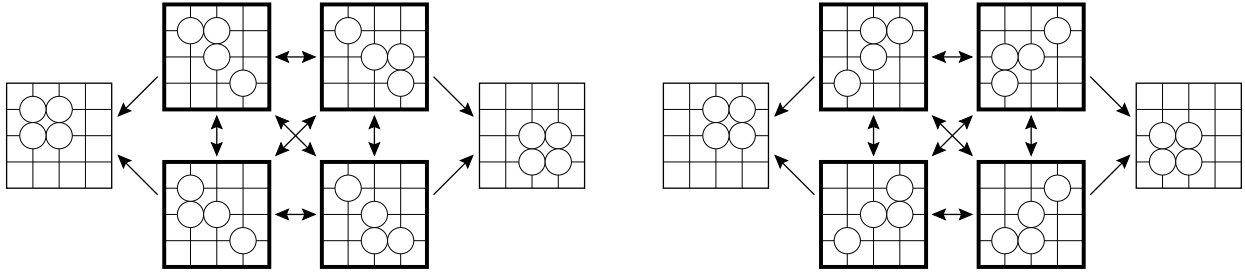}}
\caption{\label{square grid 1 extra}
  Subgraph of configuration space reachable from full-span configurations
  (outlined in bold) with one extra token.}
\end{figure}

Thus we explore the notion of \emph{extra tokens}, a set of tokens whose
removal does not reduce the span of the configuration.  Lemma \ref{span
monotone} and Figure \ref{square grid 0 extra} shows that we need at least one
extra token.  In fact, the two opposite diagonals on the square grid shown in
Figure \ref{square grid 0 extra} are difficult to reach from each other; as
shown in Figure \ref{square grid 1 extra}, even one extra token is
insufficient.  What is surprising is that a small number of extra tokens
seem to be generally sufficient to make the configuration space strongly
connected.  We prove this for the square grid in the next section.

\section{Square Grid}
\label{Square Grid}

% Our main result is about the 2-adjacency configuration space for the square
% grid.  First note that one extra coin does not suffice to maintain the span,
% as shown in Figure \ref{square grid 1 extra}.

This section analyzes coin-moving puzzles on the square grid, using the tools
from the previous section.  In particular, we show that with just two extra
coins, we can reach essentially every configuration on the square grid with the
same or smaller span.  The only restriction is that the extra coins can only
be destined for positions that are adjacent to at least two other coins.

\begin{theorem} \label{square grid}
On the square grid with the 2-adjacency restriction and unlabeled coins,
configuration $A$ can be re-arranged into configuration $B$ if there are coins
$e_1$ and $e_2$ such that $\span (A \setminus \{e_1, e_2\}) \supseteq \span (B
\setminus \{e_1, e_2\})$ and each $e_i$ is adjacent to two other coins in $B$
(excluding $e_1$ or $e_2$).
Furthermore, there is an algorithm to find such a re-arranging sequence using
$O(n^3)$ moves and $O(n^3)$ time, where $n$ is the number of coins.
\end{theorem}

%--- Old theorem statement:
% \begin{theorem} \label{square grid}
% For the square grid, any configuration can be re-arranged into any
% configuration with the same or smaller span using two extra coins.
% There is an algorithm to find such a re-arranging sequence using $O(n^4)$ moves
% and $O(n^4)$ time, where $n$ is the number of coins.
% %Let the board $G$ be the square grid, and fix the coin set $T$.  If two
% %configurations $C$ and $C'$ satisfy $\span C \supseteq \span C'$, then $C$ can
% %be re-arranged into $C'$ using two extra coins, using $O(|T|^3)$ moves.
% %----
% %On the square grid, if $A$ and $B$ are configurations with the same span, then
% %$A$ is reachable from $B$ using two extra coins.
% \end{theorem}

We prove this theorem by showing that every configuration (in particular, $A$
and $B$) can be brought to a canonical configuration with the same span via a
sequence of (mostly) reversible moves.  As a consequence, we can move from
any configuration $A$ to any other $B$ by routing through this canonical
configuration.

Our proof uses the model of picking up and dropping coins, which can be
emulated as described in Section \ref{Picking Up and Dropping Tokens}.
However, we must be careful how we pick up and drop coins, so that the
resulting moves are reversible.  For example, initially we pick up the extra
coins $e_1$ and $e_2$, and then drop them temporarily wherever needed.  For
re-arranging the source configuration $A$ into the canonical configuration,
this step may not result in reversible moves, but fortunately this is not
necessary in this case.  For re-arranging the destination configuration $B$
into the canonical configuration, however, reversibility is crucial, and is
guaranteed by the condition in the theorem of each $e_i$ being adjacent to at
least two other coins.

%--- This is covered below.
% For convenience, whenever we find a coin other than $e_1$ or $e_2$ whose
% removal does not reduce the span, we pick it up.  This rule simply removes
% clutter from the board, making it easier to argue about making moves; we can
% effectively assume that most destinations are unoccupied.  Of course, we must
% be careful when we later drop these picked-up coins, so that the resulting
% moves are reversible.  This problem will be dealt with in Section \ref{Final
% Sweep}, after all other coins are in their canonical configuration.

\subsection{Basics}
\label{Basics}

We begin with some preliminary lemmas.  A \emph{rectangle} is the full
collection of coins between two $x$ coordinates and two $y$ coordinates.  The
\emph{half-perimeter} of a rectangle is the number of distinct $x$ coordinates
plus the number of distinct $y$ coordinates over all coins in the rectangle.
The \emph{distance} between two sets of coins is the minimum distance between
two coins from different sets.

\begin{lemma}
For the square grid, the span of any configuration is a disjoint union of
(finite) rectangles with pairwise distances at least $3$.
\end{lemma}

% \begin{proof}
% Because a single coin is a rectangle, $\span A$ is certainly a disjoint
% union of rectangles.  We also want to show that each connected component
% is a rectangle, and different connected components are distance $3$ or more
% apart.

% Suppose $C$ is a connected component of $\span A$.  
% Define 
% $(a_0,b_0),(a_1,b_1)$ by
% $$a_0=min\left\{a_i|
% [a_i,b_i]\in \span A\right\},\>
% b_0=min\left\{b_i|
% [a_i,b_i]\in \span A\right\}$$
% $$a_1=max\left\{a_i|
% [a_i,b_i]\in \span A\right\},\>
% b_1=max\left\{b_i|
% [a_i,b_i]\in \span A\right\}$$
% We may assume that $a_0\not=a_1$ and $b_0\not=b_1$, since in these cases
% $\span A$ is a rectangle, and the result is easy.
% We claim that $C=R_{(a_0,b_0),(a_1,b_1)}$.  Suppose not, then there
% is a position $(a,b)\in R_{(a_0,b_0),(a_1,b_1)}\setminus C$ with 
% one of the positions adjacent to $(a,b)$ in $C$.  We can follow round
% the boundary of $C$ from this position until we reach a position
% in $R_{(a_0,b_0),(a_1,b_1)}\setminus C$  with two neighbors in $C$.
% But then this position would also be in $C$.  Hence we have a
% contradiction, and thus the connected components are rectangles.
% Clearly any two disconnected rectangles must be at least distance
% three apart, or else a coin could be added, contradicting that we have
% added all possible coins in constructing $\span A$.
% \end{proof}

\begin{lemma} \label{min chain size}
For each rectangle (connected component) of the span,
say with half-perimeter $h$, there must be at least $\lceil h/2 \rceil$ coins
within that rectangle in the configuration.
\end{lemma}

The following beautiful proof of this lemma has been distributed among several
people, but its precise origin is unknown.  We first heard it from Martin
Farach-Colton, who heard it from Peter Winkler, who heard it from Pete Gabor
Zoltan, who learned of it through the Russian magazine \emph{Kvant}
(around 1985--1987).

\medskip

\begin{proof}
Consider how the (full) perimeter changes as we compute the span of the coins
within the rectangle.  Initially we have $n$ coins, say, so the perimeter is at
most $4n$.  Each coin that we add while computing the span satisfies the
2-adjacency restriction, so the perimeter never increases.  In the end we must
have a rectangle with perimeter $2h$.  Hence $4n \geq 2h$, i.e., $n \geq h/2$,
and because $n$ is integral, $n \geq \lceil h/2 \rceil$.
\end{proof}

\subsection{Canonical Configuration}

\begin{figure}
\centerline{\includegraphics[scale=\coinscale]{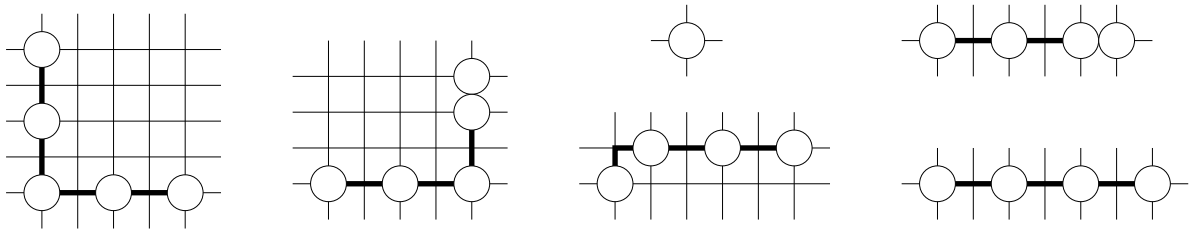}}
\caption{\label{L}
  Examples of L's.}
\end{figure}

Observe that a chain has span equal to its smallest enclosing rectangle.  We
define an \emph{L} to be a particular kind of chain, starting and ending at
opposite corners of the rectangular span, and arranged along two edges of this
rectangle, with the property that it has the minimum number of coins.  See
Figure \ref{L} for examples.  More precisely, if the half-perimeter of the
rectangle (along which the L is arranged) is $2k$, then there must be precisely
$k$ coins, every consecutive pair at distance exactly two from each other.  And
if the half-perimeter is $2k+1$, then there must be precisely $k+1$ coins,
every consecutive pair at distance exactly two from each other, except the last
pair which are distance one from each other.  In general, for half-perimeter
$h$, an L has $\lceil h/2 \rceil$ coins.

While L's can have any orientation, the \emph{canonical L} is oriented
like the letter L, starting at the top-left corner, continuing past the
lower-left corner, and ending at the bottom-right corner.

\begin{figure}
\centerline{\includegraphics[scale=\coinscale]{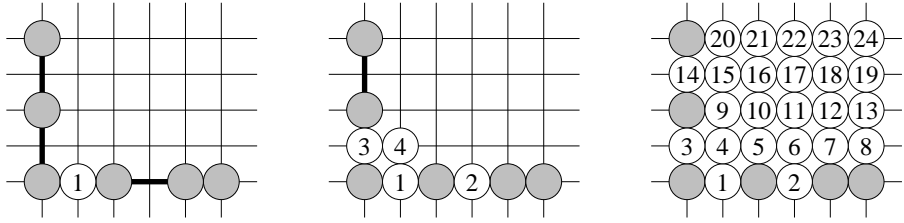}}
\caption{\label{canonical}
  Examples of the canonical configuration of $k$ coins within a
  rectangular span.  (Left) One coin in addition to the canonical L.
  (Middle) Four coins in addition to the canonical L.
  (Right) All 24 additional coins.}
\end{figure}

Given a configuration, or more precisely, given its span and the number of
coins in each connected component of the span, we define the \emph{canonical
configuration} as follows.  Refer to Figure \ref{canonical} for examples.
Within each connected component (rectangle) of the span, say with
half-perimeter $h$, we arrange the first $\lceil h/2 \rceil$ coins into the
canonical L.  (Lemma \ref{min chain size} implies that there are at least this
many coins to place.)  Any additional coins are placed one at a time, in the
leftmost bottommost unoccupied position.

This definition of the canonical configuration is fairly arbitrary, but it has
the useful property that each successive position for an additional coin is a
valid destination, given the previously placed additional coins.  This allows
us to focus on forming the canonical L, and then picking up all additional
coins and dropping them in the order shown on the right of Figure
\ref{canonical}.

\subsection{Canonicalizing Algorithm}

The main part of proving Theorem \ref{square grid} is to show an algorithm for
converting any configuration into the corresponding canonical configuration,
using a sequence of (mostly) reversible moves.  We will apply induction (or,
equivalently, recursion) on the number of coins.  That is, we assume that any
configuration with fewer coins can be re-arranged into its canonical
configuration.

For now, we assume that there are no extra coins in addition to $e_1$ and
$e_2$.  For if there were such a coin, we could immediately pick it up.  Then
we have a simpler configuration: it has one fewer coin.  Thus we can apply the
induction hypothesis, and re-arrange the remaining coins into their canonical
configuration.  Finally we must drop the previously picked-up coin in the
appropriate location.  This aspect is somewhat trickier than it may seem: if we
are not careful, we may make an irreversible move.  We delay this issue to
Section \ref{Final Sweep}.

The overall outline of the algorithm is as follows:

  \begin{enumerate}
  \item Initialize the set of L's to be one for each coin.
  \item Until the configuration is canonical:
    \begin{enumerate}
    \item Pick two L's whose bounding rectangles are distance at most two
          from each other.
    \item Re-orient the L's so that the L's themselves are distance at most
          three from each other.
    \item Merge the two L's.
    \end{enumerate}
  \end{enumerate}

Normally, each iteration of Step~2 decreases the number of L's by one, so the
algorithm would terminate in at most $n$ iterations.  However, at any time we
may find an extra coin in addition to $e_1$ and $e_2$, and pick it up.
Fortunately, this operation can only split one L into at most two L's.
Thus we can charge the cost of creating an extra L to the event of picking up
an extra coin, which can happen at most $n$ times.  Hence, the total number of
iterations of Step~2 is $O(n)$.

In the following two sections, we describe how Steps~2(b) and~2(c) can be done
in $O(n^2)$ moves each.  These bounds result in a total of $O(n^3)$ moves.  The
running time of the algorithms will be proportional to the number of moves.

\subsubsection{Re-orienting L's}
\label{Re-orienting L's}

There are eight possible orientations for an L, depending at which corner it
starts, and whether it hugs the top edge or bottom edge of the rectangular
span.  We will only be concerned with four different types of orientations,
depending on whether it looks like the letter L rotated $0$, $90^\circ$,
$180^\circ$, or $270^\circ$.  In other words, we are not concerned with the
parity issue of which corner might have two adjacent coins.

\begin{figure}
\centerline{\includegraphics[scale=\coinscale]{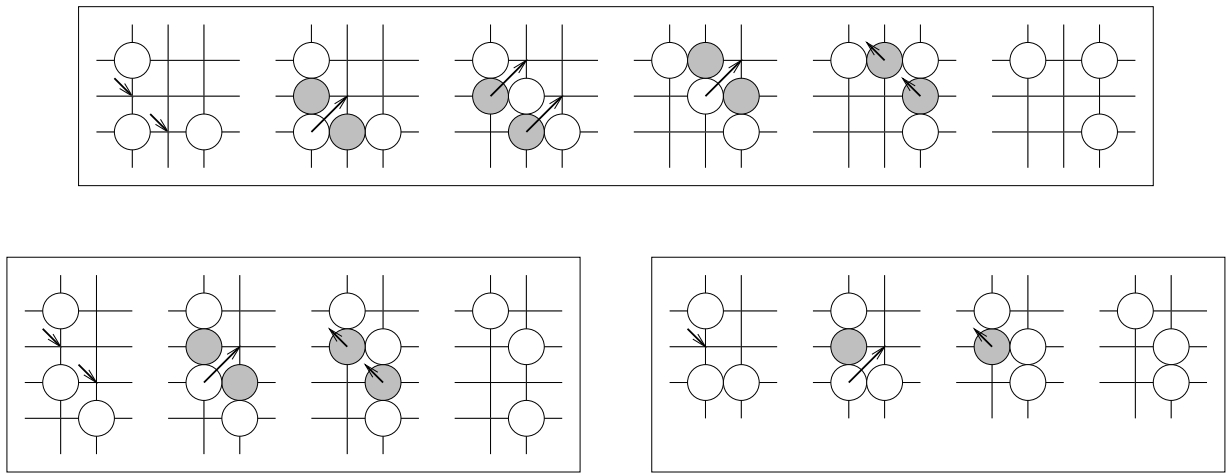}}
\caption{\label{flip 3L}
  Flipping an L consisting of 3 coins.  Extra coins are shaded.}
\end{figure}

\begin{figure}
\centerline{\includegraphics[scale=0.6]{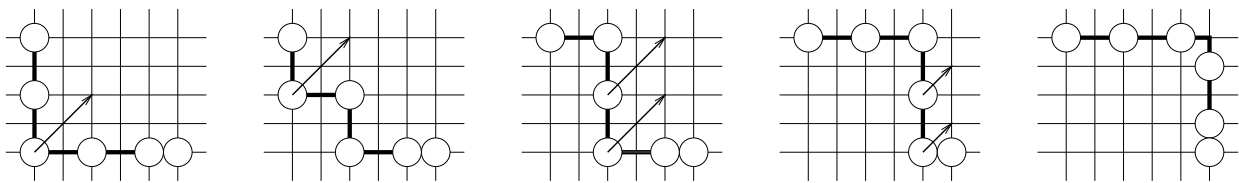}}
\caption{\label{flip L}
  Flipping a general L, using the subroutines in Figure \protect\ref{flip 3L}.}
\end{figure}

It is relatively easy to \emph{flip} an L about a diagonal, using two extra
coins.  Figure \ref{flip 3L} shows how to do this in a constant number of moves
for an L consisting of three coins.  Figure \ref{flip L} shows how to use these
subroutines to flip an L of arbitrary size.  Basically, we use the flips of
three-coin L's to ``bubble'' the kink in the L up to the top, repeatedly until
it is all the way right.  The total number of moves is $O(n^2)$, and they can
easily be computed in $O(n^2)$ time.

\begin{figure}
\centerline{\includegraphics[scale=\coinscale]{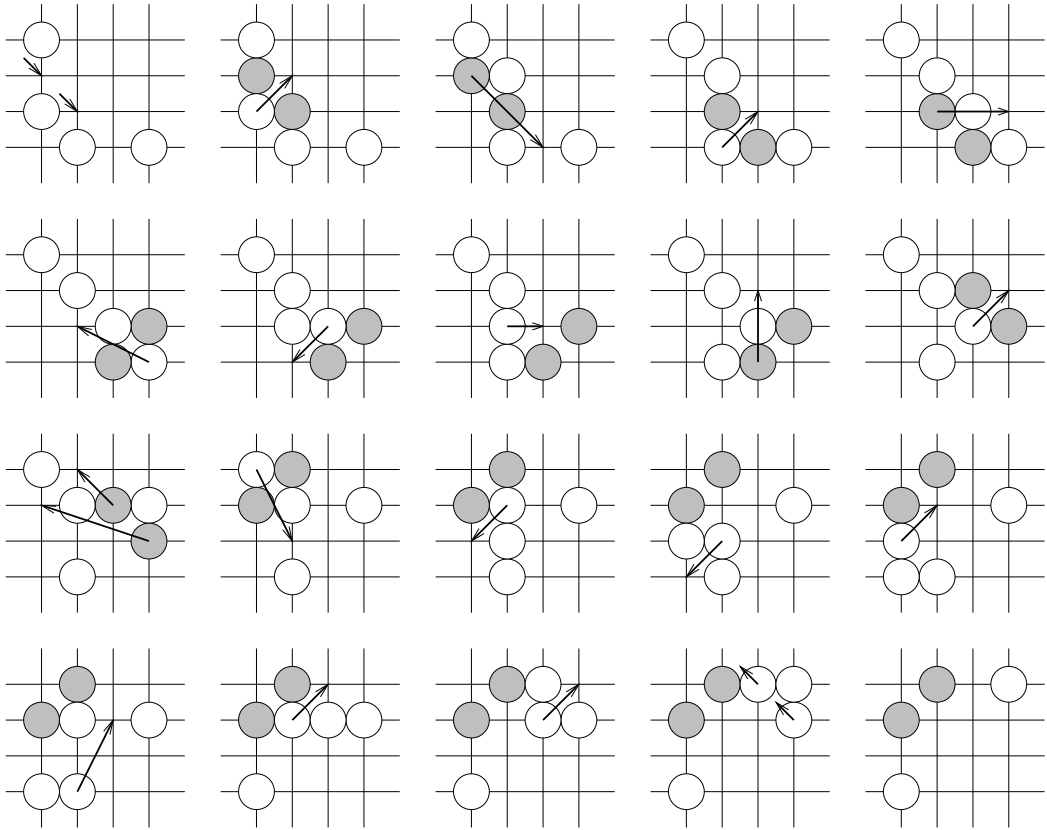}}
\caption{\label{rotate L 4x4}
  One method for rotating an L with a square span.  Although this example
  places the extra coins in the final configuration, this is not necessary.}
\end{figure}

The more difficult re-orientation to perform is a \emph{rotation} of an L by
$\pm 90^\circ$.  Perhaps one of the most surprising results of this paper is
that this operation is possible with two extra coins.  One way to do it for a
square span, shown in Figure \ref{rotate L 4x4}, is to convert the L into a
diagonal, and then convert more and more of the diagonal into a rotated L.
This is the basis for our ``diagonal-flipping'' puzzle in Figure
\ref{diagonal}.

A simpler way to argue that L's can be rotated is shown in Figure \ref{rotate
L}.  Assume without loss of generality that the initial orientation is the
canonical L.  First we apply induction to the subconfiguration of all coins
except the top-left coin.  Thus all rows except the third row contain at most
one coin each, assuming the L consists of at least three rows.  Now we apply
local operations in $3 \times 3$ or $3 \times 2$ rectangles (similar to Figure
\ref{flip 3L}) to move the top-left coin to the far right.  (We cannot perform
this left-to-right motion in one step using induction, because there may be
only three rows, and hence all coins may be involved in this motion.)  Finally
we flip the L in the top three rows, as described above, thereby obtaining the
desired result.  Again the number of moves and computation time are both
$O(n^2)$.  Note that the same approach of repeated local operations applies
when the L consists of only two rows.

\begin{figure}
\centerline{\includegraphics[scale=0.5]{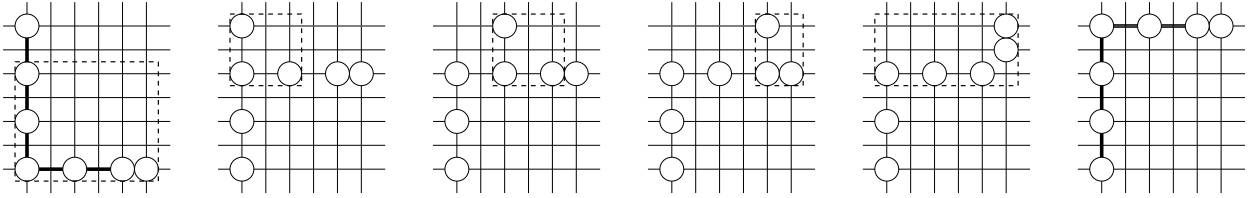}}
\caption{\label{rotate L}
  A general method for rotating an L.  The first step is to apply induction,
  and the remaining steps apply subroutines similar to
  Figure \protect\ref{flip 3L}.}
\end{figure}

\subsubsection{Merging L's}

%--- Old outline:
% route them close to each other
% if unnecessary coins, done
% else remove an end of an L that matters, apply induction
% how to add a coin back (shift it along using 3-coin subconfigurations)

Consider two L's $L_1$ and $L_2$ whose bounding rectangles $R_1$ and
$R_2$ are distance at most two from each other.  Equivalently, consider two L's
such that $\span (L_1 \cup L_2)$ has a single connected component.
This section describes how to merge $L_1$ and $L_2$ into a single L.
This step is the most complicated part of the algorithm, not because it is
difficult in any one case, but because there are many cases involved.

First suppose that the rectangles $R_1$ and $R_2$ overlap.
We claim that one of the L's, say $L_1$, can be re-oriented so that one of its
coins is contained in the other L's bounding rectangle, $R_2$.  This coin is
therefore in the span of the $L_2$, and hence redundant, so as described above
we can apply induction and finish the entire canonicalization process.

\begin{figure}
  \centering
  \includegraphics[scale=0.425]{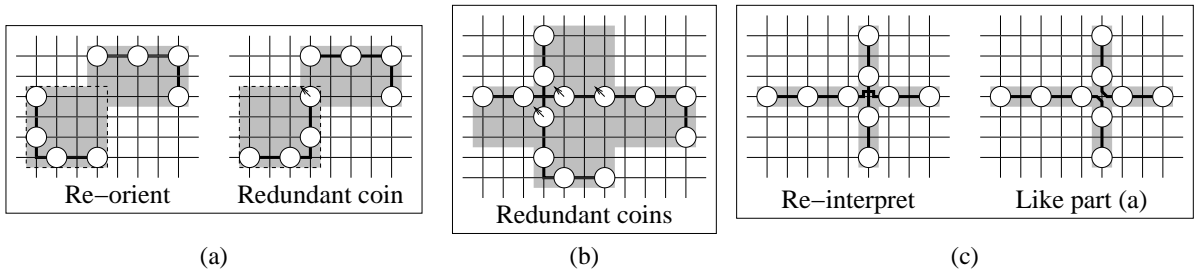}
  \caption{\label{merge L overlapping}
    Merging two L's with overlapping bounding rectangles (shaded).
    (a) The corner of one L is contained in the other L's bounding rectangle.
    (b) A ``thick plus sign'' in which at least one stroke has thickness
    more than~$1$.  (c) A plus sign in which both strokes have thickness~$1$.}
\end{figure}

To prove the claim, there are three cases; see
Figure \ref{merge L overlapping}.
If a corner of one of the bounding rectangles, say $R_1$, is in the other
bounding rectangle, $R_2$, then we can re-orient $L_1$ so that one of its end
coins is at that corner of $R_1$ and hence in $R_2$; see
Figure \ref{merge L overlapping}(a)).
Otherwise, we have rectangles that form a kind of ``thick plus sign''
(Figure \ref{merge L overlapping}(b--c));
we distinguish the two rectangles as according to whether they form the
\emph{horizontal stroke} or \emph{vertical stroke} of the plus sign.
If the vertical stroke has width at least two
(Figure \ref{merge L overlapping}(b)),
then that rectangle already contains a coin of the other L,
because that L cannot have two empty columns.
Similarly, if the horizontal stroke has height at least two,
then that rectangle already contains a coin of the other L, because that L
cannot have two empty rows.
Finally, if both strokes are of unit thickness, and there is not already a coin
in their single-position intersection (Figure \ref{merge L overlapping}(c)),
then we can splice and redefine the L's, so that one L is formed by the top
half of the vertical stroke and the left half of the horizontal stroke, and the
other L is formed by the bottom half of the vertical stroke and the right half
of the horizontal stroke, and then we have the first case in which the bounding
rectangles share a corner.

Now suppose that the rectangles $R_1$ and $R_2$ do not overlap.  Hence, either
they share no $x$ coordinates or they share no $y$ coordinates.  Assume by
symmetry that $R_1$ and $R_2$ share no $x$ coordinates.  Assume again by
symmetry that $R_1$ is to the left of $R_2$.  A \emph{leg} is a horizontal or
vertical segment/edge of an L.  Re-orient $L_1$ so that its vertical leg is on
the right side, and re-orient $L_2$ so that its vertical leg is on the left
side.  Now $L_1$ and $L_2$ have distance at most three from each other; the
distance may be as much as three because of parity.

We consider merging $L_1$ with each leg of $L_2$ one at a time.
In other words, we merge $L_1$ with the nearest leg of $L_2$ within distance
three of $L_1$, then we merge the result with the other leg of $L_2$.
The second leg can be treated in the same way as the first leg, by induction.
Thus there are two cases: either the first leg is the horizontal leg of $L_2$,
or it is the vertical leg of $L_2$.
We first show how the latter case reduces to the former case.

If the vertical leg of $L_2$ is the first leg, it can have only one coin within
the $y$ range of $R_1$ (by the assumption that extra coins are picked up).  We
can add this coin separately, as if it were a short horizontal leg of its own
L.  This leaves a portion of the vertical leg of $L_2$ outside of the $y$ range
of $R_2$.  Thus what remain of $R_1$ and $R_2$ do not share any $y$
coordinates, so we can rotate the picture $90$ degrees and return to a
horizontal problem.

\begin{figure}
\centerline{\includegraphics[scale=0.37]{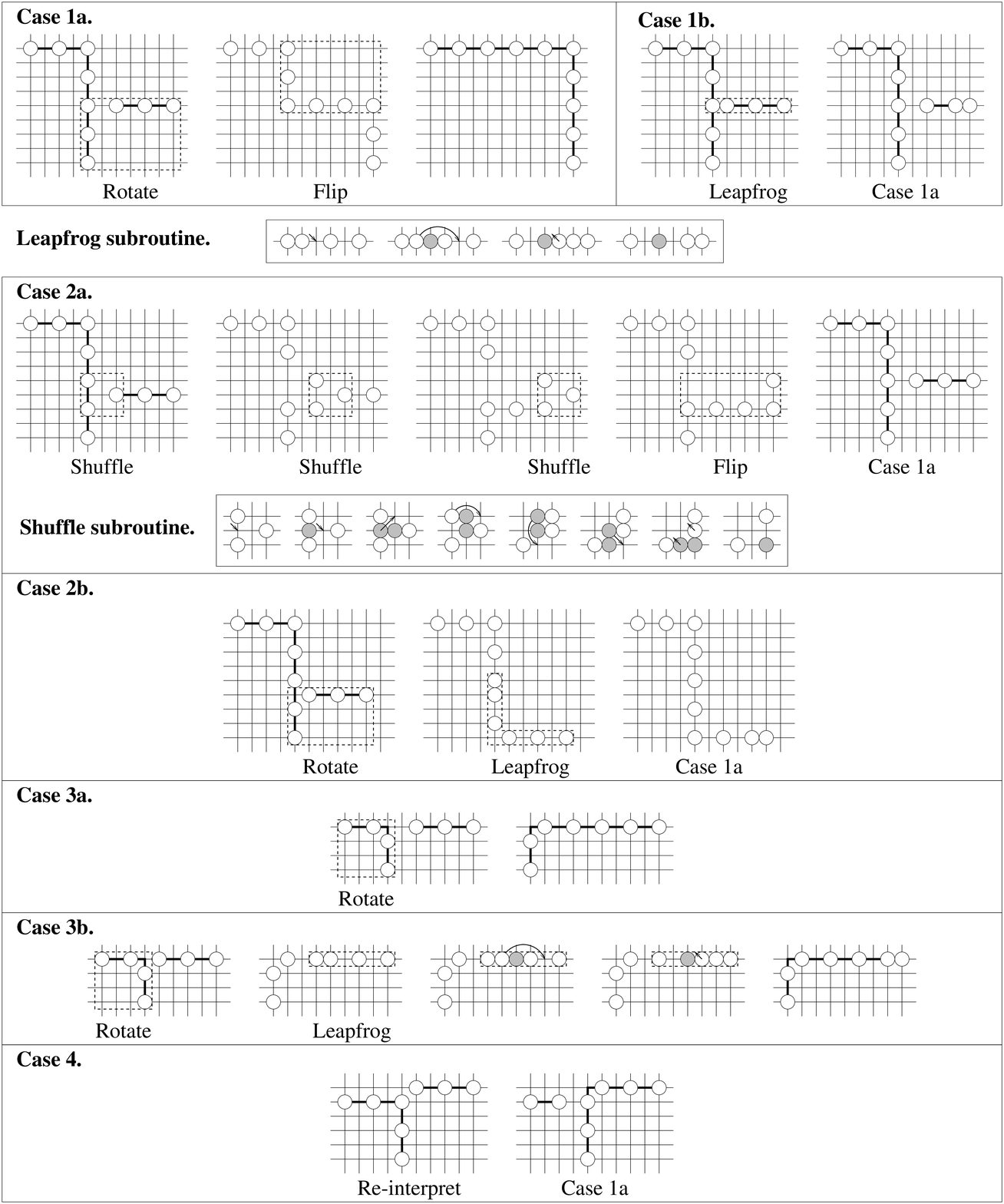}}
\caption{\label{merge L I}
  Merging an L and the horizontal leg of a nearby L.}
\end{figure}

This argument reduces merging two L's to at most three merges between an L and
a horizontal leg.  Still several cases remain, as illustrated in Figure
\ref{merge L I}.  Case 1 is when the horizontal leg is aligned with a coin in
the L.  Case 2 is when they are out of alignment.  Case 3 is a special case
occurring at the corner of the L, where the horizontal legs are aligned but
distances are higher than in Case 1.  The above three cases are subdivided into
subcases (a) and (b), depending on how close the horizontal leg is to the L.
Finally, Case 4 is when the L and horizontal leg do not share $x$ or $y$
coordinates.

By the procedures in Figure \ref{merge L I}, in all cases, the merging can be
done in $O(1)$ flips and rotations of L's, $O(1)$ leapfrogs, and $O(n)$
shuffles.  In total, $O(n^2)$ moves are required to merge an L and a horizontal
leg, or equivalently to merge two L's.

\subsection{Final Sweep}
\label{Final Sweep}

Thus far we have shown how to reversibly re-arrange a configuration ($A$ or
$B$) into the canonical configuration, using two extra coins.  However, during
this process, we may have picked up extra coins, and now need to drop them
appropriately.  In reality, these coins sit in arbitrary locations on the
board.  For re-arranging the source configuration $A$ into the canonical
configuration, the moves need not be reversible, so we can simply drop the
extra coins in the canonical order, as in Figure \ref{canonical}.  For
re-arranging the destination configuration $B$ into the canonical
configuration, we need to effectively drop these coins by a sequence of reverse
moves.

More directly, starting from the canonical configuration, we need to show how
to distribute the extra coins to arbitrary locations on the board.  We
can achieve this effect by making a complete sweep over the board.  More
precisely, we flip the L as in Section \ref{Re-orienting L's}, which has the
effect of passing over every position on the board with the operations shown in
Figure \ref{flip 3L}.  During this process, we will pass over the extra coins;
at this point we treat them as if they were picked up, applying the emulation
in Section \ref{Picking Up and Dropping Tokens}.
Then we flip the L back to its original orientation.
On the way back, whenever we apply an operation in Figure \ref{flip 3L} and
pass over the desired destination $d$ for one of the extra coins, we move the
extra coin to $d$ while there are at least two adjacent coins from the L.
By monotonicity of the flipping process, this extra coin will not be passed
over later by the flip, so once an extra coin is placed in its desired
location, it remains there.

\subsection{Reducing Span}

Now that we know any configuration can be brought to the corresponding
canonical configuration with a sequence of (mostly) reversible moves, it
follows immediately that any configuration can be re-arranged into any
configuration with the same span.  More generally, if we are given
configurations $A$ and $B$ satisfying $\span A \supseteq \span B$, we can first
pick up all coins in $A \setminus \span B$, then reversibly re-arrange both
configurations into the same canonical configuration.  Putting these two
sequence of moves together, we obtain a re-arrangement from $A \setminus \span
B$ to $B$ with some coins missing.  Then we simply drop the previously picked
up coins in the appropriate positions to create $B$.

Note that these moves need not be reversible, because we are only concerned
with the direction from $A$ to $B$.  Indeed, the moves cannot be made
reversible, because the span cannot increase (Lemma \ref{span monotone}).

This concludes the proof of Theorem \ref{square grid}.

\subsection{Lower Bound}
\label{Lower Bound}

\begin{figure}
\centerline{\includegraphics[scale=\coinscale]{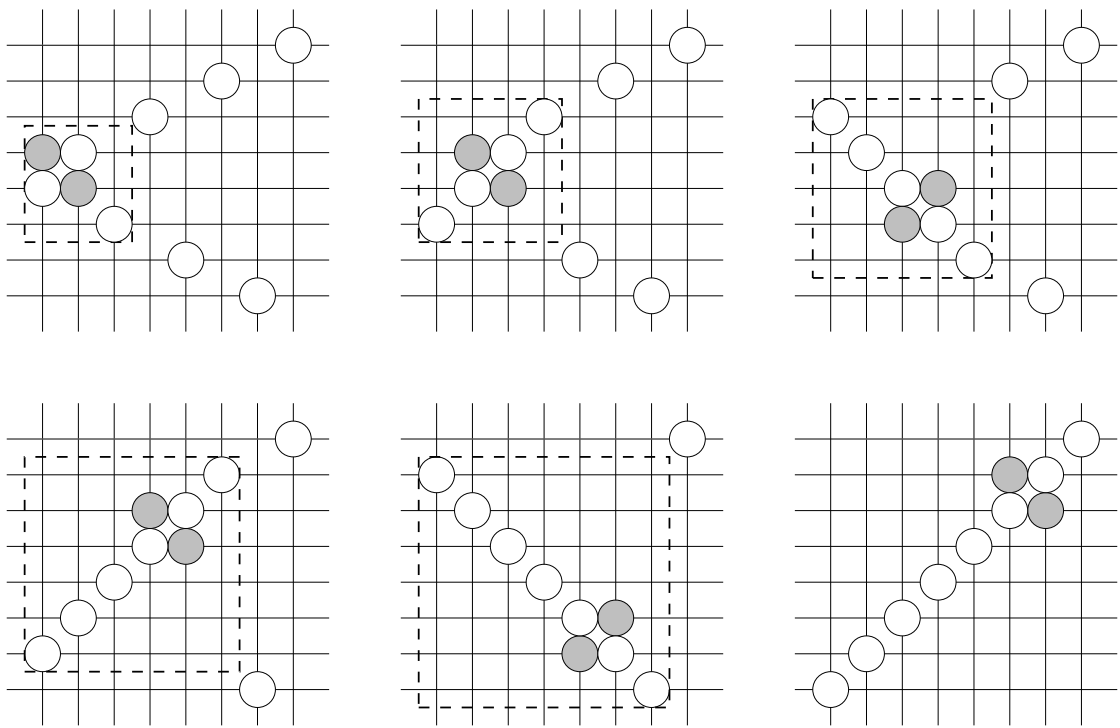}}
\caption{\label{lower bound sqr}
  Re-arranging the V-shape in the upper left into the diagonal in the
  lower right requires repeated rotations
  of diagonals as in Figure \protect\ref{diagonal} (or repeated rotations of
  L's).}
\end{figure}

The bound on the number of moves in Theorem \ref{square grid} is in fact tight:

\begin{theorem}
The ``V to diagonal'' puzzle in Figure \ref{lower bound sqr}
requires $\Theta(n^3)$ moves to solve.
\end{theorem}

\begin{proof}
We claim that re-arranging the V shape into a diagonal effectively requires
repeated ``diagonal flipping.''  At any time, only one component of coins can
be actively manipulated (drawn with dotted lines in the figure); all other
coins are isolated from movement.  Thus we must repeatedly re-arrange the
active component so that it can reach the nearest isolated coin.  More
specifically, we must re-arrange the active component into a chain starting at
the corner of the bounding rectangle that is near the isolated coin, and ending
at the opposite corner of the bounding rectangle of the active component.
These two corners alternate for each isolated coin we pick up, and that is the
sense in which we must ``flip a diagonal.''  It is fairly easy to see that
each diagonal flipping of a chain with $k$ coins takes $\Omega(k^2)$ time.
In total, the puzzle requires $\Theta(\sum_{k=1}^n k^2) = \Theta(n^3)$ moves.
\end{proof}

This theorem is the motivation for the puzzle in Figure \ref{vee}.

\subsection{Labeled Coins}
\label{Labeled Coins}

\begin{figure}
\centerline{\includegraphics[scale=0.6]{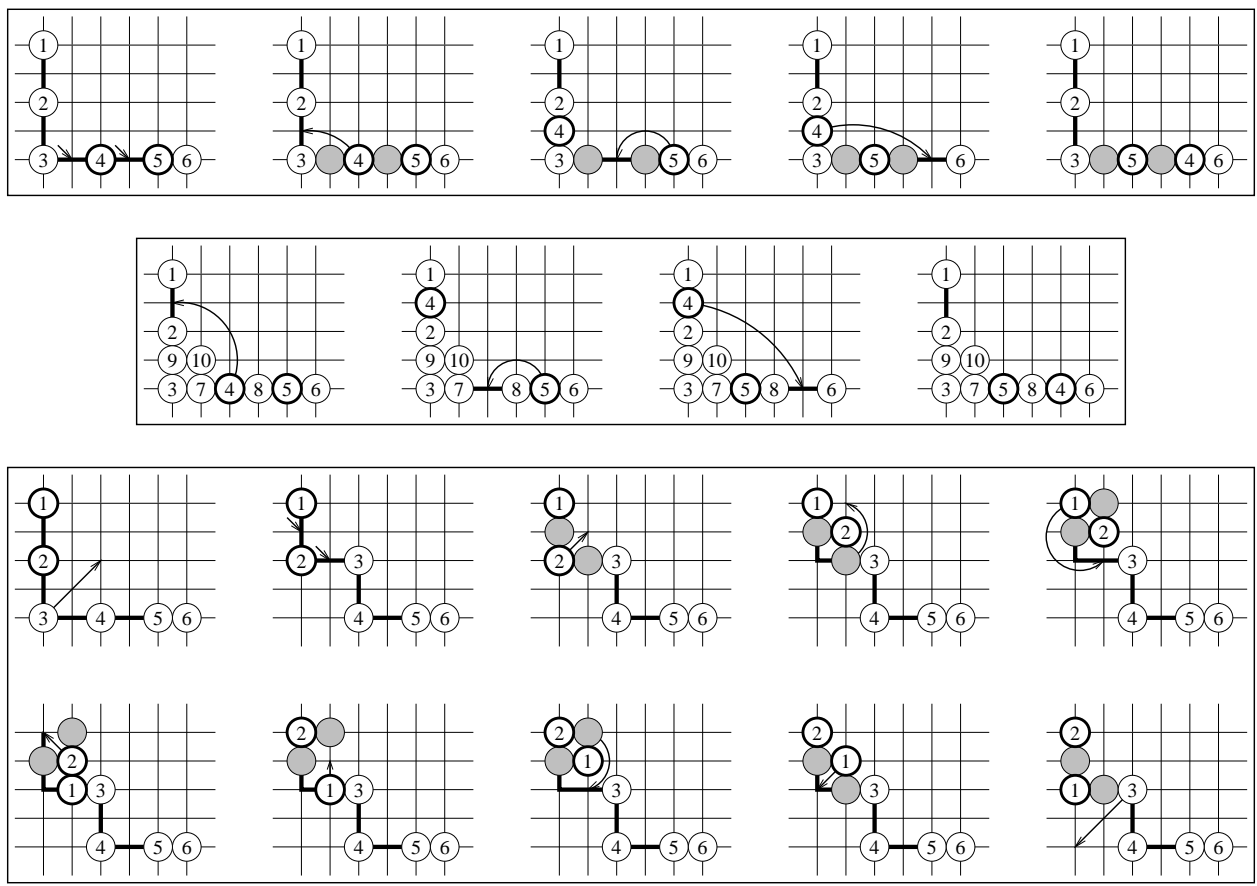}}
\caption{\label{swap L}
  Three cases of swapping coins in the canonical configuration.
  The coins to be swapped have a thick outline.}
\end{figure}

We conjecture that Theorem \ref{square grid} holds even when coins are labeled,
subject to a few constraints.  The idea is that permutation of the coins is
relatively easy once we reach the canonical configuration.  Examples of methods
for swapping coins within one L are shown in Figure \ref{swap L}.  The top
figure shows how to swap a pair of coins when the canonical configuration is
nothing more than a canonical L.  The middle figure shows how to perform the
same swap when there are four additional coins.  Note that swapping the corner
coin 3 works in exactly the same way; indeed, this method works whenever the
coins to be swapped have two other coins adjacent to them, and there is another
valid destination.  The bottom figure shows how to swap one of the end coins,
which is more difficult.  This last method begins with moving the bend of the L
toward the end coin, and then works locally on the coins $1,2,3$.

One obvious constraint for these methods is that if there are no valid moves,
permutation is impossible.  Also, if the bounding rectangle of an L has width
or height $1$, then the two end coins of the L cannot be moved.  Subject to
these constraints, Figure \ref{swap L} proves that the coins in a single
connected component of the span, other than the extra coins $e_1$ and $e_2$,
can be permuted arbitrarily.

\begin{figure}
\centerline{\includegraphics[scale=\coinscale]{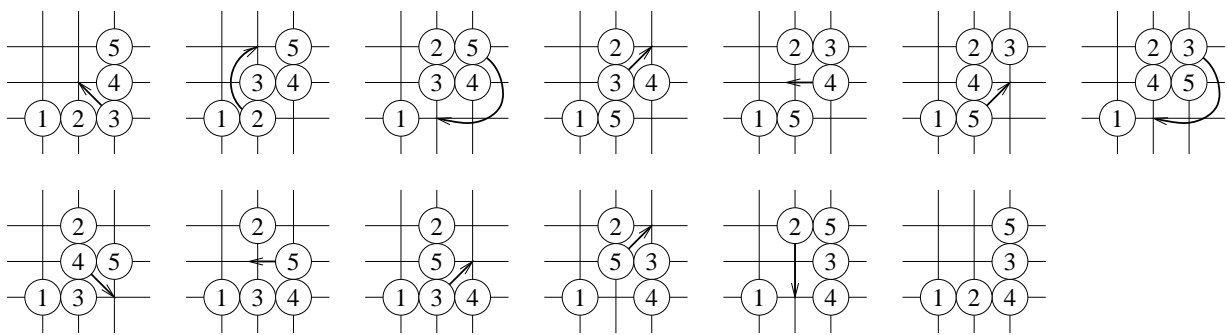}}
\caption{\label{swap5 solution}
  Solution to the puzzle in Figure \protect\ref{swap5}.}
\end{figure}

It only remains to show that a coin can be swapped with $e_1$ or $e_2$, which
implies that coins between different connected components of the span can be
swapped.  We have not proved this in general yet, but one illustrating example
is the puzzle in Figure \ref{swap5}, whose solution is shown in Figure
\ref{swap5 solution}.  The idea is that coins 2 and 4 are $e_1$ and $e_2$, and
so we succeed in swapping $e_2$ with coin 3.  A slight generalization of this
approach may complete a solution to the labeled coins.

\subsection{Fewer Extra Coins}
\label{Fewer Extra Coins}

We have shown that the configuration space is essentially strongly connected
provided there is a pair of extra coins, i.e., the removal of these two coins
does not reduce the span.  This section summarizes what we know about
configurations without this property.

If we have a span-minimal configuration with no extra coins, Lemma
\ref{span-minimal fragile} tells us that every move decreases the span.  With
an overhead of a factor of $n^2$, we can simply try all possible moves, in each
case obtaining a configuration with smaller span, which furthermore must have
an extra coin (the moved coin).  Now we only need to recursively check these
configurations.

\begin{figure}
\centerline{\includegraphics[scale=\coinscale]{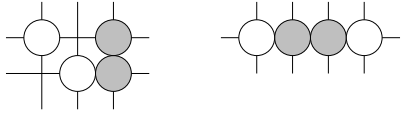}}
\caption{\label{two extra}
  The shaded coins are individually extra, but do not form a pair of
  extra coins suitable for Theorem \protect\ref{square grid}.}
\end{figure}

Unfortunately, the situation is trickier with one extra coin.  The key
difficulty is that multiple coins could individually be considered extra,
but no pair of coins is extra.  In other words, there may be two coins such
that removing either one does not reduce the span, but removing both of them
reduces the span.  Two simple examples are shown in Figure \ref{two extra}.

\begin{figure}
\centerline{\includegraphics[scale=\coinscale]{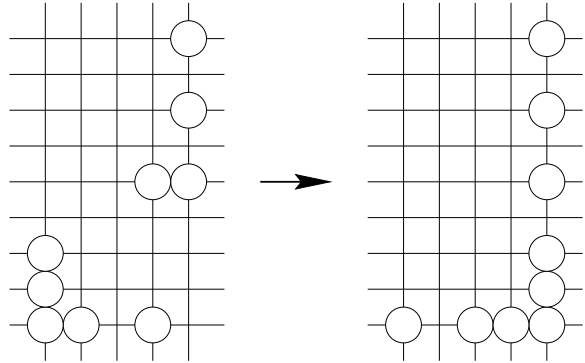}}
\caption{\label{one extra}
  A puzzle on the square grid with no initial pair of extra coins.}
\end{figure}

This difficulty makes ``one'' extra coin surprisingly powerful.  For example,
using one extra coin, an L with odd parity can be flipped, although it cannot
be rotated, and an L with even parity cannot be flipped or rotated.
In Figure \ref{one extra} we exploit this property to make an interesting
solvable puzzle initially with no pair of extra coins; it takes significant
work before a pair of extra coins appears.

\section{Conclusion}
\label{Conclusion}

We have begun the study of deciding solvability of coin-moving puzzles and more
generally token-moving puzzles.  We gave an exact characterization of solvable
puzzles with labeled coins on the equilateral-triangle grid.  By introducing
the notion of a constant number of extra coins, we have given a tight theorem
characterizing solvable puzzles on the square grid.  Specifically, we have
shown that any configuration can be re-arranged into any configuration with the
same or smaller span using two extra coins, and that this is best possible in
general.  The number of moves is also best possible in the worst case.

Several open questions remain:

  \begin{enumerate}
  \item What is the complexity of solving a puzzle using the fewest moves?
        %--- This is flat-out wrong:
        %Even the exact combinatorics of turning a general pyramid upside-down
        %as in Figure \ref{upside-down} are open \cite{Gardner-1975-penny}.
  \item How do our results change if moves are forced to be \emph{slides}
        that avoid other coins?  We conjecture that Theorem \ref{triangular
        grid} still holds for unlabeled coins.
  %\item What puzzles on the square grid with zero or one extra coins are
  %      solvable?  Can this be decided in polynomial time?
  %\item Can our results about square grids be generalized to labeled
  %      coins?  We outline some approaches to solving this in
  %      Section \ref{Labeled Coins}.
  \item Can we extend our results on the square grid to the hypercube lattice
        in any dimension?
  \item Can we combine Theorems \ref{triangular grid} and \ref{square grid}
        to deal with a mix of the square and equilateral-triangle lattice,
        like the second puzzle in Figure \ref{HOH}?
  \item Can we prove similar results for general graphs?
  \end{enumerate}

\section*{Acknowledgments}

We thank J.\ P.\ Grossman for writing a program to find optimal solutions to
the puzzles in Figures \ref{diagonal} and \ref{spindle}.

\bibliography{coinsliding}
\bibliographystyle{plain}

\end{document}